# On interpretation of force measurements in fluids: regular and thermal forces

J. Tóthová[1], L. Glod[2], G.Vasziová[1], V. Lisý[1]

[1]Department of Physics, Technical University of Košice,
Park Komenského 2, 042 00 Košice, Slovakia
[2]Department of Mathematics and Physics, University of Security Management,
Kukučínova 17, 040 01 Košice, Slovakia

**Abstract**
The paper is devoted to the problem of the determination of regular and thermal forces acting on microscopic and smaller objects in fluids. One of the methods how regular forces are determined is the measurement of the drift velocity of Brownian particles. We have obtained an exact expression for this velocity within the hydrodynamic theory of the Brownian motion. It is shown that the influence of the inertial and memory effects can be significant in the force determination when the experimental times are sufficiently short. In the second part of the work, within the same theory, we study the properties of the thermal force driving the particles in incompressible fluids. We show that the usual assumption for the Kubo´s generalized Langevin equation (called the "fundamental hypothesis") that the thermal force at a time $t$ and the velocity of the particle in preceding times are uncorrelated, leads to an unexpected super-diffusion of the particle. To obtain the Einstein diffusion at long times, the mentioned hypothesis must be abandoned, which however does not contradict to causality. Finally, we consider the "color" of thermal noise, recently measured experimentally [Th. Franosch *et al.*, Nature 478, 85 (2011)], and correct the interpretation of these experiments.

## 1. Introduction

Due to the central role that the conception of force plays in physics and with the increased interest to investigations of systems at small space and time scales, the experimental methods of the determination of ultra-small forces attract great attention. At present, new methods of such measurements have been developed [1]. The principal problem in these studies on micro- and nano-objects is connected with the influence of omnipresent thermal noise. As distinct from macroscopic systems, the noise can essentially affect the motion of small objects and the forces originating from the noise can even exceed the conservative forces in the system. This can lead to a completely incorrect interpretation of the experiments. This question is not solved up today and there is an interesting discussion about it in the literature [1 - 4]. In [1, 4], two methods of force measurements have been used to determine the influence of thermal noise on a colloidal particle near a wall in the presence of gravitational field. One of the methods consisted in measuring the particle drift velocity and the subsequent use of the overdamped Langevin equation [5] to determine the full force acting on the particle. In the second method the force was determined from the known particle-wall interaction potential using the equilibrium Boltzmann distribution. The results of these methods were strongly contrasting: the obtained forces deviated both in their magnitudes and even in their sign. As a resolution of these discrepancies, criticized however in [2], the choice of the anti-Itô conception of stochastic calculus has been suggested for the case of spatially inhomogeneous diffusion. In the present contribution we address one more problem of the measurements of forces acting on microscopic and nanoscopic objects that exists even in the case of homogeneous diffusion (for unbounded particles and constant forces, or the forces linearly depending on the particle position). This problem arises with lowering the time of measurement when the particles undergo unsteady motion for which the standard Langevin description becomes inappropriate. We show how due to the hydrodynamic aftereffect the experimentally determined force, even if it is known to be constant, appears to depend on the relation between the time of measurement and a characteristic time of the loss of memory in the system.

In the second part of the work we deal with the correlation properties of the random force driving the motion of Brownian particles in fluids. It is shown that when the commonly accepted properties of this force in incompressible fluids are assumed, this results in an unexpected motion of the particles, which will be of super-diffusive character. Since normal diffusion should take place, which is an undoubted experimental fact, this contradiction needs to be resolved. We show that the expected Einstein diffusion at long times requires that the time correlation function of the random force at the time $t$ and the particle velocity at previous moments of time must be nonzero. This apparent paradox is explained. We than calculate the correlation function of the random force itself and show that its properties (the color) significantly differ from those found in the literature [6, 7].

## 2. On the regular force measurements at short time scales

It follows from the overdamped Langevin equation that when a microscopic body is suspended in a liquid, a force $F$ applied to a body results in a drift velocity $\upsilon = F/\gamma$, where $\gamma$ is the object's friction coefficient. This is true for large constant forces, if the inertial effects are neglected [1, 4]. However, when the drift force amplitude is comparable to the effect of thermal noise and the force depends on the particle position, the equation for $F$ must be corrected [1] since the measured velocities are statistically distributed. Moreover, an additional term $-\alpha\gamma(x)\mathrm{d}D(x)/\mathrm{d}x$, referred to as *spurious force*, should be added to the force $F(x) = \gamma(x)\langle\upsilon(x)\rangle$ (for the force changing in the direction $x$). Here, $D(x)$ is the position-dependent diffusion coefficient and $\alpha$ is a constant from the interval [0, 1]. For Brownian particles the preferred value is $\alpha = 1$ [5] but there is no common agreement as to this choice [2, 8, 9], which significantly affects the stochastic calculus. One more problem, to our knowledge not considered so far in the interpretations of force measurements, appears in situations when inertial effects (and, consequently, the memory in the particle dynamics) can play a role [10]. Even if the applied force is position independent, but the observation times become comparable to the characteristic time of the loss of memory in the system, the vorticity time $\tau_R = R^2\rho/\eta$, where $R$ is the particle radius and $\rho$ and $\eta$ are the density and viscosity of the solvent, the discussed method is not applicable. We will show it coming from the generalized Langevin equation [5, 6, 10],

$$M\dot{\upsilon}_t(t) + \gamma\upsilon_t(t) + \int_0^t \Gamma(t-t')\dot{\upsilon}_t(t')\mathrm{d}t' = F + \zeta(t). \tag{1}$$

Here, $\zeta(t)$ is a random noise force with zero mean driving the particles of mass $M_p$ ($M = M_p + M_s/2$, $M_s$ is the mass of the solvent displaced by the particle), and $F$ is a regular force to be determined. The force $\zeta$ is, due to the fluctuation dissipation theorem (FDT), connected to the dissipative properties of the system. We consider the realistic case when the memory in the system is of the hydrodynamic kind, *i.e.*, the resistance force against the particle motion (the Boussinesq-Basset force, here rewritten as the integral from zero to $t$, see also below Section 4) follows from the non-stationary Navier-Stokes equations of motion for incompressible fluids [11, 12]. Then the kernel $\Gamma$ is $\Gamma(t) = \gamma(\tau_R/\pi t)^{1/2}$. The usual Brownian relaxation time is connected to the Stokes friction coefficient as $\tau = M/\gamma$. Equation (1) for $F = 0$ describes the zero-mean fluctuations $\upsilon(t)$. Here we are interested in the question how the thermal noise influences the determination of the force $F$. Let $F$ be constant, as it is for a freely falling particle in a fluid. We express the velocity as $\upsilon_t = \upsilon + \upsilon^*$. The deterministic part $\upsilon^*$ (the drift velocity) obeys the averaged Eq. (1), *i.e.* the equation without the random force

$$\dot{\upsilon}^* + \frac{1}{\tau}\upsilon^* + \frac{1}{\tau}\sqrt{\frac{\tau_R}{\pi}}\int_0^t \frac{\dot{\upsilon}^*(t')}{\sqrt{t-t'}}dt' = \frac{F}{M}. \qquad (2)$$

The initial condition at $t = 0$ (when the force $F$ begins to act) is $\upsilon^*(0) = 0$. Then the Laplace-transformed Eq. (2) has the following solution for $\tilde{\upsilon}^*(s) = \mathcal{L}\{\upsilon^*(t)\}$:

$$\tilde{\upsilon}^*(s) = \frac{F}{Ms}\frac{1}{\lambda_1-\lambda_2}\left(\frac{1}{\sqrt{s}-\lambda_1} - \frac{1}{\sqrt{s}-\lambda_2}\right), \qquad (3)$$

where $\lambda_{1,2} = -\left(\tau_R^{1/2}/2\tau\right)\left(1\mp\sqrt{1-4\tau/\tau_R}\right)$ are the roots of equation $s + (\tau_R s)^{1/2}\tau^{-1} + \tau^{-1} = 0$. The inverse transform has the form [13]

$$\upsilon^*(t) = \frac{F}{M}\left\{\tau + \frac{1}{\lambda_2-\lambda_1}\sum_{i=1}^{2}(-1)^i \frac{1}{\lambda_i}\exp\left(\lambda_i^2 t\right)\operatorname{erfc}\left(-\lambda_i\sqrt{t}\right)\right\}. \qquad (4)$$

The behavior of this solution at $t \to 0$ is $\upsilon^*(t) \approx Ft/M$. At long times we have $\upsilon^*(t) \approx (F\tau/M)\left[1-(\tau_R/\pi t)^{1/2}+...\right]$. Using these formulas, the force $F$ can be determined through the measured mean velocity $\upsilon^*(t)$ at any time $t$. At long times, due to the hydrodynamic aftereffect, this velocity depends on the relation $\tau_R/t$ and very slowly approaches the limiting value $F/\gamma$. The determined force is

$$F \approx \gamma\upsilon^*\left(1+\sqrt{\tau_R/\pi t}\right), \qquad \tau_R/t \ll 1. \qquad (5)$$

The stochastic motion of the particle, for constant $\gamma$, does not influence the determination of the force, since its contribution to the drift velocity is zero. It is easy to see that our correction of the standard result for the force, $F = \gamma\upsilon^*$, can be significant. To demonstrate it, let us turn to the recent work [10]. The smallest observation times in this experiment were $\sim 10^{-8}$ s. At such times the measured force should be proportional to $\upsilon^*/t$. At longer times, when the ratio $\tau_R/t$ is small, Eq. (5) applies. For spherical particles 1µm in radius, which are suspended in water at room temperatures, we have $\tau_R \sim 10^{-6}$ s. Thus, at the times $\sim 10^{-5}$ s the correction represents almost 20% and slowly drops with the increase of time, approaching 1% of $\gamma\upsilon^*$ at $t \cong 3$ms.

## 3. Is the Brownian motion in fluids of super-diffusive character?

In the standard Langevin approach to the motion of a BP the friction force that the particle feels during its motion is the Stokes-Einstein force. When, however, the non-stationary motion of the particle in an incompressible fluid is studied, this approach fails and this resistance force should be replaced by the Boussinesq-Basset force that reflects the memory in the system or the so-called hydrodynamic aftereffect (see [11] and references therein). Such a situation can be described by the generalized Langevin equation (GLE) (1) with $F = 0$. Now our aim is to calculate the velocity autocorrelation function (VAF) of the particle, $\phi(t) = \langle\upsilon(t)\upsilon(0)\rangle$, and its mean square displacement (MSD). For the motion along the direction $x$, $X(t) = \langle \Delta x^2(t) \rangle = \langle [x(t) - x(0)]^2 \rangle$. Here $\langle \ldots \rangle$ stays for statistical averaging. Since the quantities $\upsilon(t)$ and $x(t)$ are stochastic variables, we do not use zero initial conditions, as it is often done for them [7]. Instead, assuming the initial equilibrium between the particle and the solvent, in agreement with the equipartition theorem for the particle of mass $M$, the condition $\phi(0) = k_B T/M$ is used for the VAF. Then, multiplying Eq. (1) by $\upsilon(0)$ and assuming that $\langle \zeta(t)\upsilon(0) \rangle = 0$, after the average one obtains

$$M\dot{\phi} + \gamma\phi + \int_0^t \Gamma(t-t')\dot{\phi}(t')\,dt' = 0. \tag{6}$$

In the Laplace transformation the solution for $\tilde{\phi}(s) = \mathcal{L}\{\phi(t)\}$ is

$$\tilde{\phi}(s) = \frac{k_B T}{M}\frac{M + \tilde{\Gamma}(s)}{\gamma + s\left[M + \tilde{\Gamma}(s)\right]}, \tag{7}$$

with $\tilde{\Gamma}(s) = \gamma \tau_R^{1/2} s^{-1/2}$. The inverse transform of (7) is found after expanding this expression in simple fractions. For $\lambda_{1,2}$, defined after Eq. (3), we have

$$\tilde{\phi}(s) = \frac{k_B T}{M s^{1/2}}\left(\frac{C_1}{s^{1/2} - \lambda_1} + \frac{C_2}{s^{1/2} - \lambda_2}\right), \tag{8}$$

with $C_1 = (\lambda_1 + \tau_R^{1/2}\tau^{-1})/(\lambda_1 - \lambda_2)$, and $C_2$ obtained by exchanging the roots $\lambda_1 \rightleftarrows \lambda_2$. The time $\tau_R$ is defined before Eq. (1). Using the inverse transform $\mathcal{L}^-\{s^{-1/2}(s^{1/2} - \lambda)^{-1}\} = \exp(\lambda^2 t)\operatorname{erfc}(-\lambda\sqrt{t})$ [13], where erfc is the complementary error function, we find

$$\phi(t) = (k_B T/M)\sum_{i=1}^{2} C_i \exp(\lambda_i^2 t)\operatorname{erfc}(-\lambda_i\sqrt{t}), \tag{9}$$

with the asymptotes

$$\phi(t) \approx (k_B T/M)(1 - t/\tau + \ldots), \quad t \to 0, \tag{10}$$

$$\phi(t) \approx \frac{k_B T}{M}\left(\frac{\tau_R}{\pi t}\right)^{1/2}\left[1 - \frac{\tau_R}{2t}\left(1 - \frac{2\tau}{\tau_R}\right) + \ldots\right], \quad t \to \infty. \tag{11}$$

The MSD is obtained as $X(t) = 2\int_0^t (t-s)\phi(s)\,ds$. This equation is derived representing the distance a particle moves in time as an integral of its velocity, $x(t) - x(0) = \int_0^t \upsilon(s)\,ds$:

$$X(t) = \frac{2k_B T}{M}\frac{1}{\lambda_1 - \lambda_2}\left\{\frac{\lambda_2}{\lambda_1^4}\left[\frac{4}{3\sqrt{\pi}}\left(\lambda_1 t^{1/2}\right)^3 + \lambda_1^2 t \right.\right.$$
$$\left.\left. + \frac{2\lambda_1}{\sqrt{\pi}}t^{1/2} + 1 - \exp(\lambda_1^2 t)\operatorname{erfc}(-\lambda_1 t^{1/2})\right] - (\lambda_1 \rightleftarrows \lambda_2)\right\}. \tag{12}$$

As $t \to \infty$, up to the first term decreasing with time we have

$$X(t) = \frac{2k_B T}{M}\left[\frac{4}{3}\left(\frac{\tau_R t^3}{\pi}\right)^{1/2} + (\tau - \tau_R)t + 2\left(1 - \frac{2\tau}{\tau_R}\right)\left(\frac{\tau_R^3 t}{\pi}\right)^{1/2}\right.$$

$$-\left(\tau^2 - 3\tau\tau_R + \tau_R^2\right) + \left(\tau_R - 3\tau\right)\left(\tau_R - \tau\right)\left(\frac{\tau_R}{\pi t}\right)^{1/2} + \ldots \Bigg]. \tag{13}$$

Since $X(t \to \infty) \sim t^{3/2}$, the solution has a super-diffusive character. At short times the MSD shows the ballistic behavior, $X(t \to 0) \sim k_B T t^2/M$ [10].

However, is the unexpected result (12) – (13) correct? Equation (9) and the following equations (11) – (13) significantly differ from those found in the literature for the VAF as a solution of Eq. (1) with $F = 0$ [14 - 17],

$$\phi(t) = \frac{k_B T}{M} \frac{1}{\lambda_1 - \lambda_2} \left[\lambda_1 \exp(\lambda_2^2 t)\operatorname{erfc}(-\lambda_2 \sqrt{t}) - \lambda_2 \exp(\lambda_1^2 t)\operatorname{erfc}(-\lambda_1 \sqrt{t})\right], \tag{14}$$

which looks quite similarly to Eq. (9) but has the asymptote very different from Eq. (11):

$$\phi(t) \approx \frac{k_B T}{2M} \frac{\tau \tau_R^{1/2}}{\pi^{1/2} t^{3/2}} \left[1 - \frac{3}{2}\left(1 - \frac{2\tau}{\tau_R}\right)\frac{\tau_R}{t} + \ldots\right], \qquad t \to \infty. \tag{15}$$

The MSD that follows from Eq. (14) is

$$X(t) = 2D\Bigg\{ t - 2\left(\frac{\tau_R t}{\pi}\right)^{1/2} + \tau_R - \tau$$

$$+ \frac{1}{\tau}\frac{1}{\lambda_2 - \lambda_1}\left[\frac{\exp(\lambda_2^2 t)}{\lambda_2^3}\operatorname{erfc}(-\lambda_2 \sqrt{t}) - \frac{\exp(\lambda_1^2 t)}{\lambda_1^3}\operatorname{erfc}(-\lambda_1 \sqrt{t})\right]\Bigg\}, \tag{16}$$

with the long time behavior

$$X(t) \approx 2Dt\left\{1 - 2\left(\frac{\tau_R}{\pi t}\right)^{\frac{1}{2}} + \frac{2}{9}\left(4 - \frac{M_p}{M_s}\right)\frac{\tau_R}{t} - \frac{1}{9\sqrt{\pi}}\left(7 - 4\frac{M_p}{M_s}\right)\left(\frac{\tau_R}{t}\right)^{\frac{3}{2}} + \ldots\right\}, \quad t \to \infty, \tag{17}$$

where $D = k_B T \tau / M$ is the Einstein diffusion coefficient of the particle. Only at $\tau_R \to 0$ we have the same terms $\sim t$ and the next constant term in Eqs. (12) and (17). The characteristic times $\tau_R$ and $\tau$ are not independent; they are connected by the relation $\tau_R/\tau = 9\rho/(2\rho_p + \rho)$, where $\rho_p$ is the density of the particle. Thus, Eqs. (9) - (13) describe the normal diffusion when the density of the solvent is much smaller than that of the particle.

However, the question remains how to get the full solutions (14) and (16), which are frequently repeated in the literature. For particles with $\rho_p$ close to $\rho$ the difference between our solutions and those from the literature can be significant. In all the mentioned papers (except [14]), zero initial condition for the particle velocity is assumed. For example, in Ref. [16] both the projection of the particle position vector on the axis $x$, $x(t)$, and the velocity, $\upsilon(t)$, are assumed zero at $t = 0$: $x(0) = \upsilon(0) = 0$. It is argued that within the approximation of small thermal displacements there is no loss of generality in choosing these trivial initial conditions. Similarly, one finds in Ref. [17] that the velocity of the particle is determined by its velocity at earlier times via backflow effects in the fluid, but it is assumed that the particle is at the equilibrium position $x(0) = 0$ and at rest for $t \leq 0$. The latter condition is used also in the work [15], where independently the same solution for the VAF and MSD as in [16] has been obtained in an elegant (but incorrect, see Sec. 4) way based on the linear response theory [18]. Implicitly, as will be shown below, $\upsilon(0) = 0$ also in the recent works [7]. In our opinion, the use of the conditions $x(0) = 0$ and $\upsilon(0) = 0$ is not consistent for random variables $x(t)$ and $\upsilon(t)$ (they "never" can be met simultaneously). Moreover, it is either assumed or it follows from the calculations in all the works that the equipartition theorem holds, *i.e.* that $\langle \upsilon^2(0) \rangle$ is nonzero. We will show that if the above assumptions are abandoned, the correct (normal) diffusion can be obtained only if the random force at $t > 0$ is correlated with the particle velocity at $t = 0$. That is, we must require that $\langle \zeta(t)\upsilon(0) \rangle$ is nonzero at $t > 0$. This can be shown in the next section.

## 4. Properties of thermal random force in incompressible fluids

In spite of the above mentioned inconsistencies the solutions (14) and (16) [14 - 17] correctly describe the expected Einstein diffusion at long times for any relation between the densities $\rho_p$ and $\rho$. This problem could be resolved as follows. If we assume that the studied process is in thermodynamic equilibrium, the initial value $\upsilon(0)$ (for which we assume that the equipartition holds) should be the result of the long time memory in the system. The Langevin equation, rewritten with the integral from $-\infty$ to $t$ can be given the form [11]

$$M\dot{\upsilon}(t) + \gamma\upsilon(t) + \int_{-\infty}^{t} \Gamma(t-t')\dot{\upsilon}(t')\mathrm{d}t' = \eta(t), \tag{18}$$

where $\zeta(t)$ from Eq. (1) is $\zeta(t) = \eta(t) - \int_{-\infty}^{0} \Gamma(-t')\dot{\upsilon}(t')\mathrm{d}t'$. While we have $\langle \eta(t)\upsilon(0) \rangle = 0$ due to causality, now we can assume that the force $\zeta(t)$ and the velocity $\upsilon(0)$ correlate. Let the correlator $\langle \zeta(t)\upsilon(0) \rangle = Z(t)$ with the Laplace transform $\tilde{Z}(s)$. In the same way as before, we find the equation for the VAF $\phi(t)$

$$M\dot{\phi} + \gamma\phi + \int_{0}^{t} \Gamma(t-t')\dot{\phi}(t')\mathrm{d}t' = Z(t), \tag{19}$$

and its solution in the Laplace transform,

$$\tilde{\phi}(s) = \frac{k_B T}{M} \frac{1 + (\tau_R/s)^{1/2}\tau^{-1} + \tilde{Z}(s)/k_B T}{s + (\tau_R s)^{1/2}\tau^{-1} + \tau^{-1}}. \tag{20}$$

Let us require that in the long time limit the particle is in the diffusion regime with the Einstein's MSD proportional to $t$. This may happen only if $\tilde{\phi}(s)$ at $s \to 0$ tends to a constant, namely, $\tilde{\phi}(s) \to k_B T/\gamma = k_B T\tau/M$. Consequently, at small $s$ we must have $\tilde{Z}(s)/k_B T \approx -(\tau_R/\tau^2 s)^{1/2}$. Other terms on the right that could go to zero if $s \to 0$ must be excluded since at $s \to \infty$ there should be $s\tilde{\phi}(s) \approx k_B T/M$ to have the right $t = 0$ limit. The relation $\tilde{Z}(s)/k_B T = -(\tau_R/\tau^2 s)^{1/2}$ thus holds for every $s$, so that the VAF is

$$\tilde{\phi}(s) = \frac{k_B T}{M} \frac{1}{s + (\tau_R s)^{1/2}\tau^{-1} + \tau^{-1}}, \tag{21}$$

which corresponds to the solution (14) from the literature. Note that if we only require more generally that $\phi(t) \to 0$ as $t \to \infty$, we find that at $s \to 0$ the correlator $\tilde{Z}(s)$ must behave as $\tilde{Z}(s) \approx As^{-\mu}$, $\mu < 1$, i.e., $Z(t) \approx At^{\mu-1}/\Gamma(\mu)$, where $0 < \mu < 1$ and $\Gamma$ is the Gamma function. The Einstein diffusion corresponds to $\mu = 1/2$.

Consider for a moment the GLE in its most common form [19] with the velocity in the memory integral instead of acceleration,

$$M\dot{\upsilon}(t) + \int_0^t \Gamma(t-t')\upsilon(t')\,dt' = \xi(t). \tag{22}$$

Here $M$ is the particle mass and $\xi(t)$ the random force. Multiplying this equation by $\xi(0) = M\dot{\upsilon}(0)$ and averaging, with the use of the relations for the stationary process, $\langle\dot{\upsilon}(0)\upsilon(t)\rangle = -\langle\dot{\upsilon}(t)\upsilon(0)\rangle$, $\langle\dot{\upsilon}(0)\upsilon(0)\rangle = 0$, Eq. (22) can be rewritten to

$$H(t) = \langle\xi(t)\xi(0)\rangle = -M^2 \frac{d^2}{dt^2}\langle\upsilon(t)\upsilon(0)\rangle - M\int_0^t \Gamma(t-t')\frac{d}{dt'}\langle\upsilon(t')\upsilon(0)\rangle\,dt'. \tag{23}$$

In the Laplace transform, using $\ddot{F}(t) \div s^2\tilde{F}(s) - sF(+0) - \dot{F}(+0)$ and $\dot{F}(t) \div s\tilde{F}(s) - F(+0)$, we obtain

$$\tilde{H}(s) = -M\left[s\tilde{\phi}(s) - \phi(0)\right]\left[Ms + \tilde{\Gamma}(s)\right] = M\phi(0)\tilde{\Gamma}(s). \tag{24}$$

This is the well-known relation between the memory kernel $\Gamma$ and the correlator of the random force, $H(t) = k_B T\Gamma(t)$ (the second FDT) [19].

By the same way we obtain for the correlator $N(t) = \langle\zeta(t)\zeta(0)\rangle$ in our problem of the hydrodynamic Brownian motion

$$\tilde{N}(s) = \gamma^2\tilde{\phi}(s) - \left[M^2 s + Ms\tilde{\Gamma}(s) - \gamma\tilde{\Gamma}(s)\right]\left[s\tilde{\phi}(s) - \phi(0)\right]. \tag{25}$$

The Einstein diffusion at long times is reached if Eq. (21) holds, when

$$\tilde{N}(s)/\phi(0) = \gamma M + \tilde{\Gamma}(s)[Ms - \gamma] \tag{26}$$

and

$$N(t) = k_B T\gamma\left[\delta(t) - \frac{1}{\tau}\sqrt{\frac{\tau_R}{\pi t}}\theta(t) - \frac{1}{2}\sqrt{\frac{\tau_R}{\pi t^3}}\theta(t)\right], \tag{27}$$

where $\theta(t)$ is the Heaviside function. This expression is similar to that found in Ref. [7], except the second term in square brackets that is missing there.

Now, let us consider the studied problem from another view. If we average Eq. (22) with a regular external force $F(t)$,

$$M\dot{\upsilon}(t) + \int_0^t \Gamma(t-t')\upsilon(t')\,dt' = \xi(t) + F(t), \tag{28}$$

using $\langle\upsilon(0)\rangle = \langle\xi(t)\rangle = 0$ and the Laplace transformation (denoted by the index $s$, *e.g.* the transform of $\langle\upsilon(t)\rangle$ will be $\langle\upsilon(t)\rangle_s$), we find

$$\langle\upsilon(t)\rangle_s = \frac{\langle F(s)\rangle_s}{Ms + \tilde{\Gamma}(s)}. \tag{29}$$

From here the admittance (or mobility of the particle) is [19]

$$\tilde{\mu}(s) = \frac{1}{Ms + \tilde{\Gamma}(s)}. \tag{30}$$

Together with Eq. (21) that leads to the Einstein diffusion,

$$\tilde{\phi}(s) = \frac{k_B T}{Ms + \tilde{\Gamma}(s)}, \tag{31}$$

we find

$$\tilde{\phi}(s) = k_B T \tilde{\mu}(s). \tag{32}$$

This relation is called the basic theorem or the generalization of the first FDT [19]. It also holds

$$\langle v(t) \rangle_s = \tilde{\mu}(s) \langle F(s) \rangle_s = \frac{1}{k_B T} \tilde{\phi}(s) \langle F(s) \rangle_s. \tag{33}$$

This equation is similar to that used in Ref. [15] (Eq. 3b),

$$v(t) = \frac{1}{k_B T} \int_0^t \phi(t - t') F(t') dt', \tag{34}$$

but in [15] $F$ is the random force incorrectly assumed to be zero at $t < 0$, and $v$ is the velocity ($v(0) = 0$). However, Eq. (33) is valid for the mean velocity and the external regular force $F$. When the external force is absent, the mean velocity is zero and Eq. (33) cannot be used. The derivation of the VAF in [15] is thus not correct. But the basic theorem (32) allows us to find the correlators of the random force in the Langevin equation and the velocity. If the GLE (22) is multiplied by $v(0)$ and averaged, in the Laplace tranform we obtain for the VAF:

$$\tilde{\phi}(s) = \frac{M\phi(0) + \tilde{Z}(s)}{Ms + \tilde{\Gamma}(s)}, \tag{35}$$

where the transformed correlator $Z(t) = \langle \xi(t) v(0) \rangle$ reads

$$\tilde{Z}(s) = \tilde{\phi}(s) \left[ Ms + \tilde{\Gamma}(s) \right] - M\phi(0). \tag{36}$$

If the relation $\tilde{\phi}(s) = k_B T \tilde{\mu}(s)$ holds, $Z(t) = 0$. If we require that the normal diffusion takes place at long times, and there is equipartition at $t = 0$, again $Z(t) = 0$ at $t > 0$. In the hydrodynamic Brownian motion however the condition $Z(t) = 0$ leads to super-diffusion. We have shown above for Eq. (1) that $Z(t) = \langle \zeta(t) v(0) \rangle$ must differ from zero. This is consistent with the general relation (32): for $\tilde{\phi}(s)$ we have the equation

$$\tilde{\phi}(s) = \tilde{\mu}(s) \left\{ \left[ M + \tilde{\Gamma}(s) \right] \phi(0) + \tilde{Z}(s) \right\}, \tag{37}$$

which together with (32) gives

$$\tilde{Z}(s) = -\phi(0) \tilde{\Gamma}(s). \tag{38}$$

This is the same relation that we have obtained from the requirement that Einstein diffusion takes place at long times.

Note that the "fundamental hypothesis" $Z(t) = 0$ is used in the cited works [7] also for the hydrodynamic Brownian motion. Then the derivation of the VAF in [7] is based on the solution of the Kubo's equation (22) by multiplying it with $\upsilon(0)$ and averaging. However, Eq. (22) follows from Eq. (1) only if $\upsilon(0) = 0$ (in other case divergent terms appear when the integral containing $d\upsilon(t')/dt'$ is transformed to the integral with $\upsilon(t')$). This makes the way of solution in [7] flawed.

Finally, let us return to the question of the color of thermal noise. Although the thermal force in the Langevin equation is a physical reality and should be observable [19], its "color" has been directly measured only very recently by Franosch *et al.* [6]. They trapped Brownian particles in a harmonic potential and measured the positional autocorrelation function $\langle x(t)x(0)\rangle$ from the recorded position fluctuations of the particle. It was expected that the color of the thermal noise is not white – at different times its values correlate (the mean $\langle \zeta(t)\,\zeta(t')\rangle$ is no more proportional to the delta function $\delta(t - t')$). Ignoring the particle inertia at low frequencies (long times), the trapping force dominates over friction. Then the Langevin equation (1) with the force $Kx(t)$, where $K$ is the stiffness of the harmonic potential, is reduced to $Kx(t) \approx \zeta(t)$. From here Franosch *et al.* used $\langle \zeta(t)\zeta(0)\rangle \approx K^2\langle x(t)x(0)\rangle$ and determined

$$\langle \zeta(t)\zeta(0)\rangle = -\frac{1}{2}k_B T \gamma \left(\frac{\tau_R}{\pi t^3}\right)^{1/2}. \tag{39}$$

It is evident that this is not correct since such a derivation requires that also $Kx(0) \approx \zeta(0)$ holds, which is not true. One more mistake done in Ref. [6] is of the same nature as in [7] and consists in the already discussed incorrect transformation of the hydrodynamic Langevin equation (1) to the Kubo's GLE (22). Above, see Eq. (27), a different correlator $\langle \zeta(t)\,\zeta(0)\rangle$ has been obtained. For $t > 0$ we have

$$\langle \zeta(t)\zeta(0)\rangle = -k_B T \gamma \left(\frac{\tau_R}{\pi t}\right)^{1/2}\left(\frac{1}{\tau} + \frac{1}{2t}\right). \tag{40}$$

The time correlation function of the thermal noise in incompressible fluids thus at long times approaches zero as $t^{-1/2}$ instead of $t^{-3/2}$ as found in [6].

## 5. Conclusion

In conclusion, we have obtained an exact solution for the drift velocity of a Brownian particle in an incompressible fluid under the action of a constant force, taking into account the hydrodynamic memory in the particle motion. This velocity is proportional to the applied force but depends in a complicated manner on the time of observation $t$. At short times it is proportional to $t$ and at long times it contains algebraic tails, the longest-lived of which being $\sim t^{-1/2}$. Due to this the velocity very slowly approaches the limiting value $F/\gamma$. As a consequence, the force $F$ can significantly differ from the value that would be extracted from the drift measurements neglecting the inertial effects, which is a standard assumption in the interpretation of such experiments. The presented method is equally applicable in the case of the force linearly depending on the particle position. For nonlinear forces, first the open question about the choice of convention to be used in stochastic calculus should be resolved. The second part of the work deals with the properties of the thermal noise driving the Brownian particles. Using an effective method of solving linear generalized Langevin equations [20], we show that the expected Einstein diffusion can be obtained from the hydrodynamic Langevin equation in the form (1) only in the case when the thermal force $\zeta(t)$ at $t > 0$ correlates with the velocity of the particle at the time $t = 0$. This is in contradiction with the "fundamental hypothesis" (according to which these quantities are uncorrelated) used in a number of papers dealing with the normal and generalized Langevin equation [7]. We have found the corresponding correlator $\langle \zeta(t)\upsilon(0) \rangle$. Also, we have shown that the Einstein diffusion at long times takes place when another basic theorem from the linear response theory is applied: this theorem joins the mobility of the particle and its velocity autocorrelation function. Finally, we discuss the question of the color of the thermal noise. It was claimed in the recent work [6] that this color was experimentally measured. We show that the interpretation of those interesting experiments should be corrected and calculate the time correlation function of the thermal noise within the hydrodynamic theory of the Brownian motion. The difference between our results and those found in previous works [6, 7] is significant.

**Acknowledgement**

This work was financially supported by the Agency for the Structural Funds of the EU within the project NFP 26220120033, and by the grant VEGA 1/0370/12.